\documentclass[showpacs,preprintnumbers,twocolumn]{revtex4}
\usepackage[dvips]{epsfig}
\usepackage{psfrag}
\usepackage{graphicx}
\usepackage{slashbox}
\usepackage{amssymb, amsmath}

\begin{document}

\preprint{SU-ITP-06-11}
\preprint{NORDITA-2006-7}
\preprint{RH-09-2006}

\title{Global geometry of two-dimensional charged black holes} 
\date{April 5, 2006}

\author{Andrei V. Frolov}
\email{afrolov@stanford.edu}
\affiliation{KIPAC/SITP, Stanford University, Stanford, 
CA 94305-4060, USA}
\author{Kristj\'an R. Kristj\'ansson}
\email{kristk@nordita.dk}
\affiliation{NORDITA, Blegdamsvej 17, 2100 Copenhagen, Denmark}
\author{L\'arus Thorlacius}
\email{lth@hi.is}
\affiliation{University of Iceland, Science Institute, Dunhaga 3, 
107 Reykjav\'{\i}k, Iceland}

\begin{abstract}
The semiclassical geometry of charged black holes is studied 
in the context of a two-dimensional dilaton gravity model where 
effects due to pair-creation of charged particles can be included 
in a systematic way. The classical mass-inflation instability of the 
Cauchy horizon is amplified and we find that gravitational collapse 
of charged matter results in a spacelike singularity that precludes
any extension of the spacetime geometry. 
At the classical level, a static solution describing an eternal black
hole has timelike singularities and multiple asymptotic regions. 
The corresponding semiclassical solution, on the other hand, has
a spacelike singularity and a Penrose diagram like that of an
electrically neutral black hole. Extremal black holes are destabilized
by pair-creation of charged particles. There is a maximally charged 
solution for a given black hole mass but the corresponding geometry
is not extremal. Our numerical data
exhibits critical behavior at the threshold for black hole formation.

\end{abstract}

\maketitle

\section{Introduction}

\begin{figure}
\includegraphics[width=4cm]{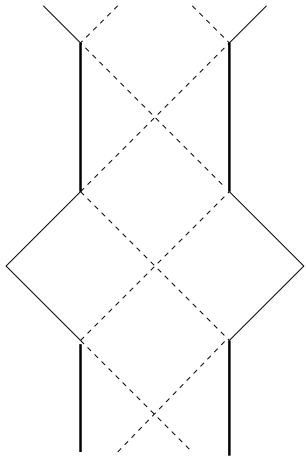}
\caption{The Penrose diagram of a classical $1+1$-dimensional 
charged black hole is the same as for a Reissner-Nordstr\"om 
black hole in $3+1$ dimensions.  The thick lines represent the 
timelike singularities and the dashed lines are the horizons. 
The structure repeats itself in the vertical direction.}
\label{fig:RN}
\end{figure}

\begin{figure}
\includegraphics[width=5.5cm]{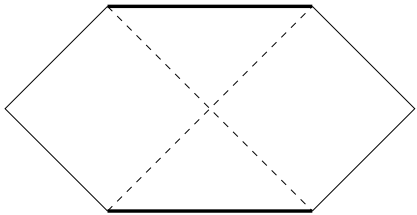}
\caption{The Penrose diagram for a semiclassical charged black hole
in 1+1 dimensions is the same as that of a 3+1-dimensional 
Schwarzschild black hole.}
\label{fig:Schwarzschild}
\end{figure}

In a recent paper \cite{Frolov:2005ps} we introduced a 
two-dimensional model for the study of quantum effects in a 
charged black hole spacetime.
The main advantage of this model is that the back-reaction on the 
geometry, due to the pair production of charged particles, can be taken
into account in a systematic way.  In the weakly coupled asymptotic region
the back-reaction amounts to a minor modification of the classical theory but 
the effect on the interior geometry of a charged black hole is more dramatic.
The timelike singularities and Cauchy horizons of a static classical charged black
hole are replaced by a simpler causal structure with a spacelike singularity
inside a single horizon. In other words, the Reissner-Nordstr\"om like Penrose 
diagram of the classical geometry, shown in Figure~\ref{fig:RN}, is replaced at 
the semiclassical level by the Schwarzschild type Penrose diagram, 
shown in Figure~\ref{fig:Schwarzschild}. This conclusion is reached by a 
combination of analytic and numerical calculations. 

The back-reaction effect on dynamical black holes formed in
gravitational collapse of charged matter is equally dramatic.
In numerical simulations based on our semiclassical equations
a spacelike singularity forms inside a single apparent horizon,
as was advocated in pioneering work of Novikov and Starobinsky
\cite{Novikov:1980ni}. This spacelike singularity replaces
the relatively weak mass inflation singularity that develops
at a null Cauchy horizon in the classical theory
\cite{Poisson:1990eh,Ori:1991,Brady:1995ni,Hod:1998gy,
Dafermos:2003wr}.

In the present paper we carry out a more detailed study of our 
two-dimensional model, elaborating on and going beyond the
results reported in  \cite{Frolov:2005ps}. In Section \ref{sec:classical} we 
discuss classical black hole solutions of  two-dimensional dilaton 
gravity coupled to an abelian gauge field.  Like four-dimensional 
Reissner-Nordstr\"om black holes these static geometries have 
timelike curvature singularities inside Cauchy horizons and the 
maximally extended spacetime contains multiple asymptotic regions 
\cite{McGuigan:1991qp,Frolov:1992xx}, as shown in Figure~\ref{fig:RN}.  
 
In Section \ref{sec:coupling} we add charged matter to the model
in order to study dynamical solutions involving gravitational 
collapse. Our choice of matter sector, i.e.\ charged Dirac
fermions, is particularly convenient for studying semiclassical
corrections to the geometry due to matter quantum effects.
Bosonization of the fermions has the combined advantage of 
including the effect of fermion pair-production at a semiclassical
level and converting the matter equations of motion into a scalar
field equation, which is more amenable to analytic and numerical 
study than the original fermion theory.
 
The resulting semiclassical equations are obtained in Section
\ref{sec:semiclassical} and we study their static solutions in 
some detail in Section \ref{sec:static}, paying attention both
to the black hole region, where we find a spacelike singularity,
and to the exterior region, where there is an outgoing flux of
charged particles due to pair-production in the electric field of
the black hole.  

We also consider maximally charged solutions for a given black 
hole mass and contrast their Schwarzschild like geometry against 
that of extremal black holes in the classical theory.

In Section \ref{sec:dynamical} we turn to the study of gravitational
collapse in the semiclassical model. We describe a leap-frog 
algorithm that is well adapted to problems of this kind and 
present numerical results that show the formation of a spacelike
singularity inside a single apparent horizon. 

Finally, we consider gravitational collapse in the 
limit of vanishing black hole mass and observe a form of
Choptuik scaling  \cite{Choptuik:1992jv}. 

\section{Classical theory}
\label{sec:classical}

Let us begin by describing classical black hole solutions of 
1+1-dimensional dilaton gravity coupled to an abelian gauge field.
The classical action is given by
\begin{equation}
\label{classact}
S_ {dg}  = \int d^2x \sqrt{-g}e^{-2\phi}
\left[R+4(\nabla\phi)^2+4\lambda^2-\frac{1}{4} F^2\right].
\end{equation}
The overall factor of $e^{-2\phi}$ in front tells us that the strength of 
both the gravitational coupling and the gauge coupling is governed 
by the dilaton field. 

This 1+1-dimensional theory can be obtained by spherical reduction 
of 3+1-dimensional dilaton gravity in the background of an extremal 
magnetically charged black
hole \cite{Callan:1992rs,Giddings:1992kn,Banks:1992ba}. 
In what follows we are mostly interested in the 1+1-dimensional 
theory in its own right as a simplified model of gravity but the 
higher-dimensional interpretation sheds light on some aspects of
the physics. 

The action (\ref{classact}) inherits a mass scale $\lambda$ from 
the 3+1-dimensional theory, which is proportional to the inverse 
of the magnetic charge of the extremal dilaton black hole.  
In the following we use units where $\lambda=1$. 
The area of the transverse two-sphere in the Einstein frame 
in 3+1 dimensions is proportional to $\psi\equiv e^{-2\phi}$, and 
hence we refer to $\psi$ as the {\it area function}. 

In order to study the formation of charged black holes in our
1+1-dimensional world, we have to add some form of charged matter 
to the theory. The detailed form of the matter action is not needed for 
this preliminary discussion and will be specified later on.

The classical equations of motion are
\begin{eqnarray}
\label{classeq1}
\frac{1}{4}R + \nabla^2\phi-(\nabla\phi)^2+1
&=& \frac{1}{16}F_{\mu\nu}F^{\mu\nu} , \\
\nabla_\mu\nabla_\nu\phi+g_{\mu\nu}((\nabla\phi)^2-\nabla^2\phi -1)
&=& \nonumber \\
\label{classeq2}
\frac{1}{4}(F_{\mu\lambda}F_\nu^{\ \lambda}-\frac{1}{4}g_{\mu\nu}
F_{\lambda\sigma}F^{\lambda\sigma})
&+&\frac{e^{2\phi}}{2} T^m_{\mu\nu} , \\
\nabla_\nu(e^{-2\phi}F^{\nu\mu})
&=& j^\mu , 
\label{classeq3}
\end{eqnarray}
where $j^\mu$ and $T^m_{\mu\nu}$ are components of the matter 
current and energy-momentum tensor, whose form depends 
on the matter system in question. The vacuum equations, with 
$j^\mu=T^m_{\mu\nu}=0$, have a two-parameter family of static 
solutions
\begin{eqnarray}
\label{bhone}
\phi  &=& -x , \\
\label{lindilmetric}
ds^2
&=& -a(x)dt^2+\frac{1}{a(x)}dx^2  ,\\
F_{tx}  &=& Q e^{-2x} ,
\end{eqnarray}
where 
\begin{equation}
a(x)=1-Me^{-2x}+\frac{1}{8}Q^2e^{-4x}.
\end{equation} 
In this coordinate system the dilaton field depends linearly on the 
spatial coordinate. The metric approaches the two-dimensional 
Minkowski metric and the coupling strength $e^\phi$ goes to zero
in the asymptotic region $x\rightarrow\infty$. The electromagnetic 
field $F_{tx}$ also goes to zero asymptotically. Its 3+1-dimensional
origins are reflected in the fact that it goes as the inverse of the
transverse area. 

Horizons occur at zeroes of the metric function $a(x)$. Thus the 
nature of the solution in the interior region depends on the 
constants $M$ and $Q$, the mass and charge of the geometry.  
Just as in the 3+1-dimensional Reissner-Nordstr\"om solution, 
there are three cases to consider for a given $\vert Q\vert >0$:
\begin{itemize}
\item
$M>\vert Q \vert/\sqrt{2}$: A charged black hole with 
two separate horizons,
\item
$M=\vert Q \vert/\sqrt{2}$: An extremal black hole where the two 
horizons coincide,
\item
$M<\vert Q \vert /\sqrt{2}$: A naked singularity.
\end{itemize}
We will focus on black holes with $M\geq\vert Q \vert/\sqrt{2}$. 
In this case $a(x)$ has two zeroes where the area function $\psi$ 
takes the values
\begin{equation}
\psi_\pm=\frac{1}{2}\left(M\pm \sqrt{M^2-\frac{1}{2}Q^2} \right) .
\label{psipm}
\end{equation}
This relation can equivalently be written
\begin{align}
M &= \psi_+ + \psi_- , \\  
Q^2 &= 8\psi_+\psi_- .
\label{psipm2}
\end{align}

The metric (\ref{lindilmetric}) is singular at $\psi=\psi_+$ but, 
since the spacetime curvature is finite there, this signals the 
breakdown of the linear dilaton coordinate system rather than a 
problem with the geometry itself. It is straightforward to find 
new coordinates which describe the solution in the interior region 
where $\psi<\psi_+$. A standard Kruskal-like extension results in 
the Penrose diagram in Figure \ref{fig:RN}, which is identical 
to that of a 3+1-dimensional 
Reissner-Nordstr\"om black hole. We will not work out that 
extension here but rather use variables that turn out to be 
convenient when we generalize our equations to include 
semiclassical effects.

In two dimensions we can write $F^{\mu\nu}=f\varepsilon^{\mu\nu}$, 
where $f$ is a scalar field and $\varepsilon^{\mu\nu}$ is an 
antisymmetric tensor, related to the Levi-Civita tensor density by
\begin{equation}
\varepsilon^{\mu\nu}=\frac{\epsilon^{\mu\nu}}{\sqrt{-g}}.
\end{equation}
We work in conformal gauge 
\begin{equation}
ds^2=e^{2\rho}(-dt^2+d\sigma^2)
\end{equation}
and look for static solutions with $j^\mu=T^m_{\mu\nu}=0$.
The classical equations reduce to 
\begin{eqnarray}
\phi''-2\rho'\phi'&=& 0 ,\\
\phi''-\rho'' +\frac{1}{2}f^2e^{2\rho} &=& 0 , \\
\phi'\rho'-\phi'^2+(1-\frac{1}{8}f^2)e^{2\rho} &=& 0 , \\
(fe^{-2\phi})' &=& 0 ,
\label{maxw}
\end{eqnarray}
where prime denotes $\frac{d}{d\sigma}$. The Maxwell equation 
(\ref{maxw}) allows us to eliminate the gauge field in favor of 
the area function, 
\begin{equation}
f = \frac{Q}{\psi} ,
\label{psiforf}
\end{equation}
with the black hole charge $Q$ appearing as an integration constant.
The electric field at the event horizon $f\vert_H\equiv f_+$ is 
given by 
\begin{equation}
f_+ =\frac{Q}{\psi_+} = \sqrt{\frac{8\psi_-}{\psi_+} },
\label{fonhorizon}
\end{equation}
where we have used $Q^2=8\psi_+\psi_-$.
Similarly, the electric field at the inner horizon is
\begin{equation}
f_- =\frac{Q}{\psi_-} = \sqrt{\frac{8\psi_+}{\psi_-} } .
\end{equation}
We note that the field at either horizon does not depend on the two 
black hole parameters independently but only on their ratio $Q/M$. 
Furthermore, the field at the inner horizon is bounded from
below, $f_- > \sqrt{8}$, for all classical black holes in
this model.

Now introduce $\xi= e^{2(\rho-\phi)}$ and define a new spatial 
coordinate $y$ via $dy=\xi d\sigma$. The remaining classical 
equations take a particularly simple form when expressed in 
terms of $\psi$ and $\xi$,
\begin{align}
\label{classicalpsi}
\ddot{\psi} &= 0,  \\
\label{classicalxi}
\ddot{\xi} &= \frac{8\psi_+\psi_-}{\psi^3}, \\
\label{classicalpsixi}
\dot{\xi}\dot{\psi} &= 4(1-\frac{\psi_+\psi_-}{\psi^2}),
\end{align}
where the dot denotes $\frac{d}{dy}$. For a charged black hole 
these equations are valid outside the outer horizon and inside 
the inner horizon. In the region between the two horizons the 
$y$ coordinate is timelike and the left hand sides of equations 
(\ref{classicalpsi})-(\ref{classicalpsixi}) change sign.

\begin{figure}
\begin{tabular}{c@{\hspace{0.2cm}}c}
\includegraphics[height=4cm]{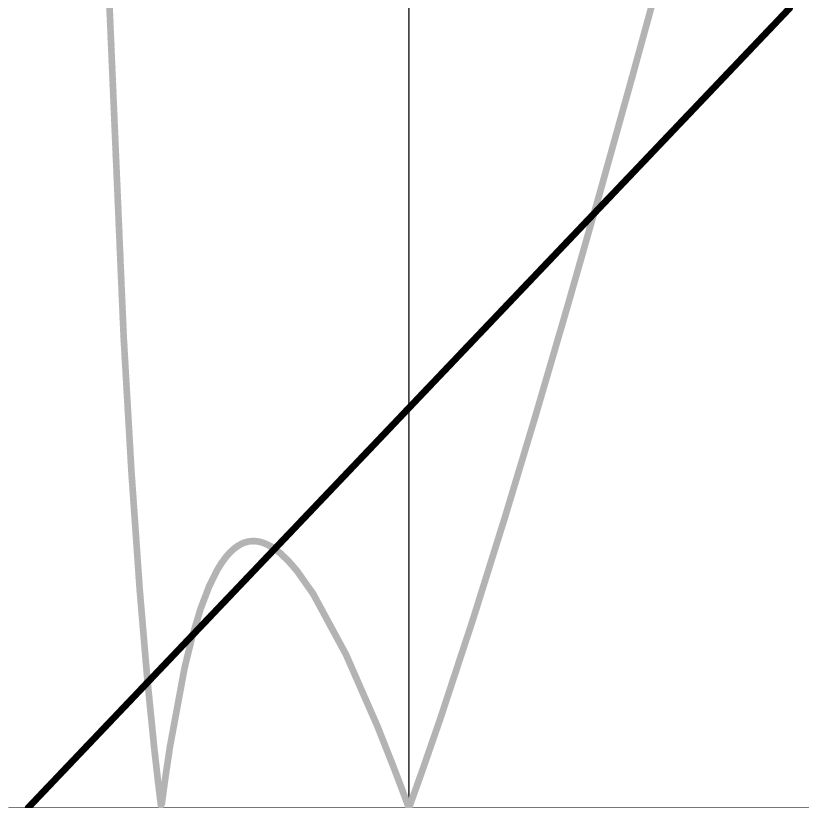} & 
\includegraphics[height=4cm]{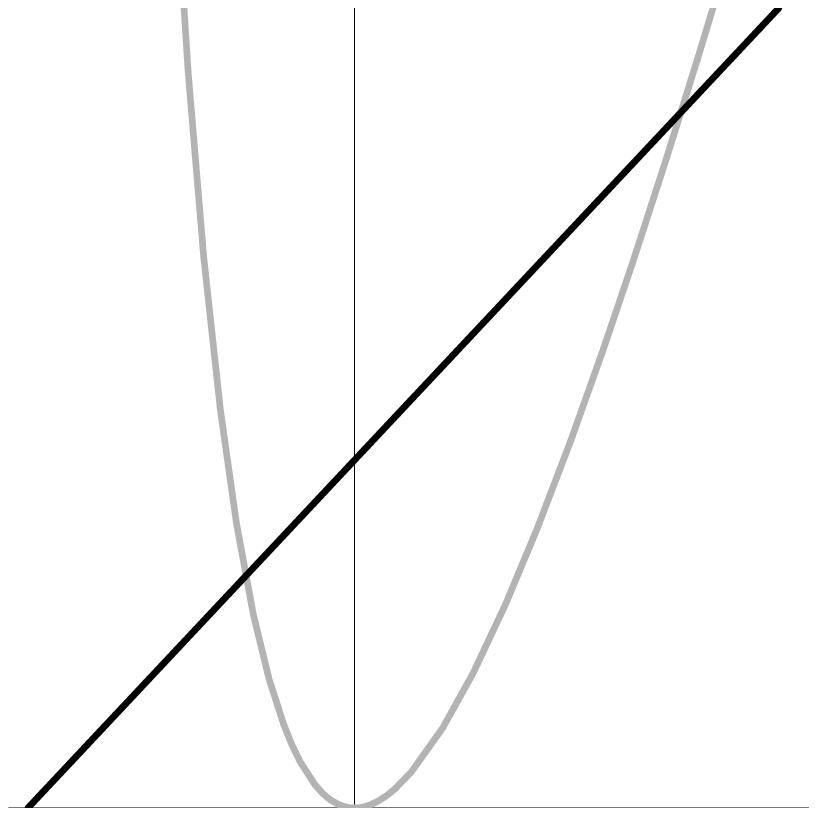} \\
(a) & (b)\\
\end{tabular}
\caption{(a) $\psi$ and $\xi$ plotted as a function of $y$ for a 
classical black hole solution. The two horizons are at the zeroes 
of $\xi$ and the curvature singularity is where $\psi$ goes to 
zero. (b)~In the extremal limit $\xi$ has a double zero at $y=0$ 
and the two horizons coincide.}
\label{fig:classical}
\end{figure}

The solution for a charged black hole, shown in 
Figure~\ref{fig:classical}a, is given by
\begin{eqnarray}
\label{classpsi}
\psi(y)
&=& \psi_+ + \alpha y, \\
\xi(y)
&=& \frac{4}{\alpha^2}
\left\vert\alpha y
+\frac{\psi_+\psi_-}{\psi_++\alpha y}
-\psi_-\right\vert ,
\label{classxi}
\end{eqnarray}
where $\alpha>0$ sets the scale of the $y$ coordinate and we have 
placed the origin $y=0$ at the outer horizon. The absolute value sign
accommodates the sign-flip in equations 
(\ref{classicalpsi})-(\ref{classicalpsixi}) in the region between the two
horizons. In the asymptotic region $y\rightarrow \infty$ the conformal 
factor of the metric approaches a constant value 
$e^{2\rho}\rightarrow 4/\alpha^2$ 
and the spacetime curvature goes to zero. If we require the metric to 
approach the standard Minkowski metric the scale parameter is fixed 
at $\alpha=2$. It turns out to be convenient, however, to allow for 
general $\alpha$ in the classical solution when comparing to 
semiclassical results.

The area function goes to zero at $y=-\psi_+/\alpha$. This is a 
curvature singularity and the solution cannot be extended any further.

In the extremal limit we have $\psi_+=\psi_-=M/2$ and the above 
solution reduces to 
\begin{eqnarray}
\psi(y)
&=& \frac{M}{2} + \alpha y, \\
\label{classextremal}
\xi(y)
&=& \frac{8y^2}{M + 2\alpha y}.
\end{eqnarray}
In this case there is a double horizon at $y=0$ and a curvature 
singularity at $y = -M/2\alpha $.  The electric field at the 
double horizon of an extremal black hole is $f_H = \sqrt{8}$ for 
all values of $M$.  The extremal solution is shown in 
Figure~\ref{fig:classical}(b).

In the limit of neutral black hole we instead have $\psi_+=M$,
$\psi_-= 0$ and the classical solution 
takes the following simple form,
\begin{equation}
\psi(y) = M + \alpha y, \quad 
\xi(y) = \frac{4}{\alpha} |y|.
\end{equation}
In this case there is only one horizon and the curvature
singularity at $y= -M/\alpha$ is spacelike.

So far, we have only considered static classical solutions.
In order to study charged black hole formation one
would couple the dilaton gravity and gauge field to some
form of charged matter and look for solutions of the 
classical equations of motion (\ref{classeq1})--(\ref{classeq3}) 
with incoming matter energy and current. As far as we know,
no closed form dynamical solutions to these equations exist.
This is perhaps not surprising given that the system is known
to exhibit intricate dynamical behavior, including a 
two-dimensional version of mass inflation 
\cite{Balbinot:1994ee,Chan:1994tb,Droz:1994aj}.
In the following section we couple the theory to charged 
matter but we do not pursue the problem of classical 
gravitational collapse. Instead we go on to include quantum
effects in the form of pair-production of charged particles
and then study the resulting semiclassical equations.

\section{Coupling to charged matter}
\label{sec:coupling}

In order to study the effect of Schwinger pair-production we add 
matter in the form of a 1+1-dimensional Dirac fermion 
$\Psi=\left(\begin{array}{c} \psi_L \\  \psi_R \end{array} \right)$ 
to the theory,
\begin{equation}
\label{matteract}
S_ {m}  = \int d^2x \sqrt{-g}\left[
i\bar\Psi\gamma^\mu ({\cal D}_\mu+ieA_\mu )\Psi 
- m\bar\Psi\Psi \right]
\end{equation}
where $e$ and $m$ are the charge and mass of our `electrons', and 
${\cal D}_\mu = \partial_\mu + 
\frac{i}{2}J_{ab}\omega^{ab}_{\phantom{ab}\mu}$ 
denotes a covariant derivative acting on 1+1-dimensional spinors.
With this matter sector, our model can be viewed as a generalization
to include gravitational effects in the `linear dilaton 
electrodynamics' developed in 
\cite{Susskind:1992gd,Peet:1992yd,Alford:1992ef}.
The current and energy-momentum carried by the fermions are given by 
\begin{eqnarray}
\label{current}
j^\mu
&=& e\bar\Psi\gamma^\mu\Psi \\
T_{\mu\nu}
&=& \frac{i}{4} \bar\Psi(\gamma_\mu {\cal D}_\nu + \gamma_\nu 
{\cal D}_\mu)\Psi \nonumber\\
&-&\frac{1}{2} g_{\mu\nu} (i\bar\Psi\gamma^\lambda
({\cal D}_{\lambda} + ieA_\lambda)\Psi - m\bar\Psi\Psi )
\label{fermionemt}
\end{eqnarray}
We could in principle look for dynamical solutions of the combined 
fermion and dilaton gravity system that describe classical black 
hole formation by an incoming flux of fermions. Our main interest 
is, however, in semiclassical geometries, with the back-reaction 
due to Schwinger pair-creation taken into account and this 
requires a different approach.

The key to including pair-creation 
is provided by the quantum equivalence between fermions 
and bosons in 1+1 dimensions. The massive Schwinger model, 
{\it i.e.} quantum electrodynamics of a massive Dirac particle in 
1+1 dimensions, is equivalent to a bosonic theory with a 
Sine-Gordon interaction \cite{Coleman:1975pw,Coleman:1976uz}.
In flat spacetime the identification between the fermion field and 
composite operators of a real boson field $Z$ is well known 
\cite{Coleman:1975pw,Coleman:1976uz}. 
This identification carries over to curved spacetime, with
appropriate replacement of derivatives by covariant derivatives, 
as long as the gravitational field is slowly varying on the 
microscopic length scale of the matter system. The description 
in terms of bosons will break down in regions where the curvature 
gets large on microscopic length scales, {\it i.e.} near curvature 
singularities, but in such regions any classical or semiclassical 
description will be inadequate anyway.

In terms of the bosonic field the matter current is given by
\begin{equation}
\label{bosoncurrent}
j^\mu =  
\frac{e}{\sqrt{\pi}}\varepsilon^{\mu\nu}\nabla_\nu Z ,
\end{equation}
and the covariant effective action for the boson is
\begin{equation}
\label{bosonact}
S_ {b}  = \int d^2x \sqrt{-g}\left[-\frac{1}{2} (\nabla Z)^2
- V(Z) -\frac{e}{\sqrt{4\pi}} \varepsilon^{\mu\nu}F_{\mu\nu} 
Z\right],
\end{equation}
where $V(Z)=c\,e\,m (1-\cos(\sqrt{4\pi}Z))$, with $c$ a numerical 
constant whose precise value does not affect our considerations. 
In order to model real electrons our fermions should have a large
charge-to-mass ratio. In this case the fermion system is highly
quantum mechanical but, since the fermion-boson equivalence
in 1+1 dimensions is an example of a strong/weak coupling 
duality, the boson system is classical precisely when
$m\ll e$. For simplicity, we set $m=0$ in most of what follows, 
but our numerical results below include runs with $m>0$.

\section{Semiclassical equations}
\label{sec:semiclassical}

We obtain the semiclassical geometry of a two-dimensional black 
hole by solving the equations of motion of the combined boson and 
dilaton gravity system, (\ref{classact}) and (\ref{bosonact}). 
We consider both dynamical collapse solutions and static solutions 
that describe eternal black holes. For the dynamical solutions it 
is convenient to work in conformal gauge with null coordinates 
$ds^2=-e^{2\rho}d\sigma^+d\sigma^-$.  
Writing $F^{\mu\nu}=f\varepsilon^{\mu\nu}$ as before, the Maxwell
equation (\ref{classeq3}), including the bosonized matter current 
(\ref{bosoncurrent}), becomes 
\begin{equation}
\label{maxwell}
\partial_\pm (e^{-2\phi}f+\frac{e}{\sqrt{\pi}}Z) =0 .
\end{equation}
Once again the gauge field can be eliminated,
\begin{equation}
\label{zforgauge}
f = \frac{1}{\psi}\left(-\frac{e}{\sqrt{\pi}}Z+q \right).
\end{equation}
By comparing to the classical result (\ref{psiforf}) we see that 
the value of the bosonic field $Z$ at a given spatial location 
determines the amount of electric charge to the left of that 
location, or `inside' it from the 3+1-dimensional point of view.  
There is also an integration constant $q$ that represents a 
background charge located at the strong coupling end of the
one-dimensional space.  In the following we will set $q=0$. 
This is natural when we consider gravitational collapse of 
charged matter into an initial vacuum configuration. 
Furthermore, if we assume that the background charge $q$ is an 
integer multiple of the fundamental charge $e$ carried by our 
fermions, then it can be set to zero by a symmetry of the 
semiclassical equations under a discrete shift of $Z$ by 
$\sqrt{\pi}$ times an integer.  For $m=0$ the shift symmetry 
becomes continuous and in that case an arbitrary background 
charge, and not just integer multiples of $e$, can be absorbed
by a shift of $Z$. For convenience we adopt units in which $e=1$
in the remainder of this paper.

We now insert expression (\ref{zforgauge}) for the gauge field,
with $q=0$ and $e=1$, into the remaining semiclassical equations.
They are somewhat simplified if we introduce $\theta=2(\rho-\phi)$
and work with the area function $\psi$ instead of the dilaton field
itself.  The resulting system of equations is 
\begin{eqnarray}
\label{psieq}
-4\partial_+\partial_-\psi
&=& \left(4-\frac{Z^2}{2\pi\psi^2}\right)e^\theta
-\frac{V(Z)e^\theta}{\psi} , \\
\label{thetaeq}
-4\partial_+\partial_-\theta
&=& \frac{Z^2e^\theta}{\pi\psi^3}
+\frac{V(Z)e^\theta}{\psi^2}, \\
\label{zeq}
-4\partial_+\partial_-Z
&=& \frac{Ze^\theta}{\pi\psi^2} +
\frac{V'(Z)e^\theta}{\psi} ,
\end{eqnarray}
along with two constraints
\begin{equation}
\label{constraints}
\partial_\pm^2\psi-\partial_\pm\theta\partial_\pm\psi 
= -\frac{1}{2}(\partial_\pm Z)^2.
\end{equation}

Equations (\ref{psieq})--(\ref{constraints}) can be solved 
numerically with initial data that describes charged matter 
undergoing gravitational collapse. In section \ref{sec:dynamical} we 
employ a finite difference algorithm to study this process but 
let us first consider the simpler problem of static solutions of 
the semiclassical equations that describe eternal black holes.

\section{Static black holes}
\label{sec:static}

For static configurations of the semiclassical equations
(\ref{psieq})--(\ref{constraints}) we require the fields to 
depend only on the spatial variable 
$\sigma= \frac{1}{2}(\sigma^+ - \sigma^-)$.
To study such solutions we proceed as in the classical theory, 
writing $\xi=e^\theta$ and defining a new spatial coordinate via 
$dy = \xi d\sigma$. Outside the event horizon the semiclassical 
equations (\ref{psieq})--(\ref{thetaeq}) reduce to 
\begin{eqnarray}
\label{staticpsieq}
\xi \ddot{\psi} +\dot{\xi}\dot{\psi}
&=& 4-\frac{Z^2}{2\pi\psi^2}
-\frac{V(Z)}{\psi} , \\
\label{staticxieq}
\ddot{\xi}
&=& \frac{Z^2}{\pi\psi^3}
+\frac{V(Z)}{\psi^2},\\
\label{staticzeq}
\xi \ddot{Z} + \dot{\xi}\dot{Z}
&=& \frac{Z}{\pi\psi^2} 
+ \frac{V'(Z)}{\psi} ,
\end{eqnarray}
where the dots once again denote derivatives with respect to $y$.
The constraints  (\ref{constraints}) become 
\begin{equation}
\label{staticconstraint}
\ddot{\psi}+\frac{1}{2}(\dot{Z})^2=0.
\end{equation}
At first sight, it appears that we have four equations for only
three fields, but the equations are not independent. Inserting 
(\ref{staticconstraint}) into (\ref{staticpsieq}) gives
\begin{equation}
-\frac{1}{2}\xi \dot{Z}^2 +\dot{\xi}\dot{\psi}
= 4-\frac{Z^2}{2\pi\psi^2}- \frac{V(Z)}{\psi},
\end{equation}
which is easily seen to be a first integral of equations
(\ref{staticxieq})--(\ref{staticconstraint}).

Inside the event horizon $y$ becomes timelike which means that
the derivative terms in equations 
(\ref{staticpsieq})--(\ref{staticzeq})
change sign in that region. 
\begin{eqnarray}
\label{staticpsieqinside}
-\xi \ddot{\psi} -\dot{\xi}\dot{\psi}
&=& 4-\frac{Z^2}{2\pi\psi^2}
-\frac{V(Z)}{\psi} , \\
\label{staticxieqinside}
-\ddot{\xi}
&=& \frac{Z^2}{\pi\psi^3}
+\frac{V(Z)}{\psi^2},\\
\label{staticzeqinside}
-\xi \ddot{Z} - \dot{\xi}\dot{Z}
&=& \frac{Z}{\pi\psi^2} 
+ \frac{V'(Z)}{\psi} ,
\end{eqnarray}
The semiclassical equations are more complicated than the 
classical ones and explicit analytic solutions are not available.
They can, however, be integrated numerically to obtain 
information about the semiclassical geometry of charged
black holes. 

\subsection{Numerical black hole solutions}

In order to find a numerical black hole solution which is well
behaved at the event horizon we start our integration at 
$y=0$ and set $\xi(0)=0$. Different black holes are parametrised
by the initial values $Z(0)$ and $\psi(0)$. Second order
equations also require initial values for first derivatives
of the fields. The choice of $\dot \xi(0)$ does not affect the 
geometry but instead sets the scale of the spatial coordinate
$y$. The remaining two initial values are provided by the
equations themselves when we impose the condition
that the solution be regular at the event horizon 
\cite{Birnir:1992by}. By setting $\xi(0)=0$ in equations 
(\ref{staticpsieq}) and (\ref{staticzeq}) for the exterior
solution (equations (\ref{staticpsieqinside}) and
(\ref{staticzeqinside}) for the interior solution)
while requiring that $\ddot \psi$ and $\ddot Z$ are 
finite at $y=0$, we obtain
\begin{eqnarray}
\label{initialdatapsi}
\dot \psi(0) &=& \frac{1}{\vert\dot \xi(0)\vert}\left(
4-\frac{Z(0)^2}{2\pi\psi(0)^2 } 
- \frac{V(Z(0))}{\psi(0)}\right), \\
\label{initialdataz}
\dot Z(0) &=& \frac{1}{\vert\dot \xi(0)\vert}\left(
\frac{Z(0)}{\pi\psi(0)^2 } 
+ \frac{V'(Z(0))}{\psi(0)}\right).
\end{eqnarray}
The exterior solution is found by starting with these
initial values at the black hole horizon at $y=0$ for some
$\dot\xi(0)>0$ and integrating equations 
(\ref{staticpsieq})-(\ref{staticzeq}) numerically towards $y>0$. 
For the integration into the black hole we instead 
use equations (\ref{staticpsieqinside})-(\ref{staticzeqinside}) and
change the sign of $\dot\xi(0)$. 

\begin{figure}
\begin{tabular}{c@{\hspace{0.8cm}}c}
\includegraphics[height=4.80cm]{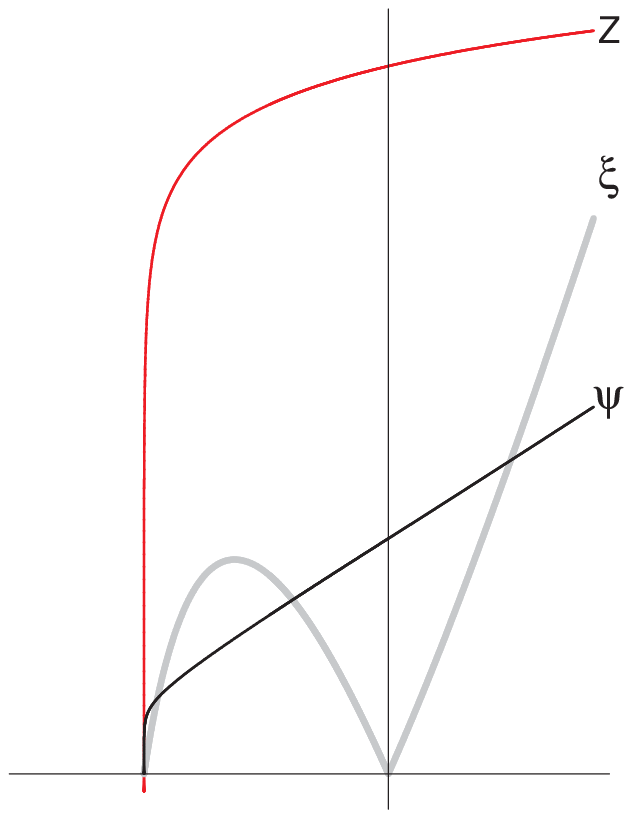} &
\includegraphics[height=4.80cm]{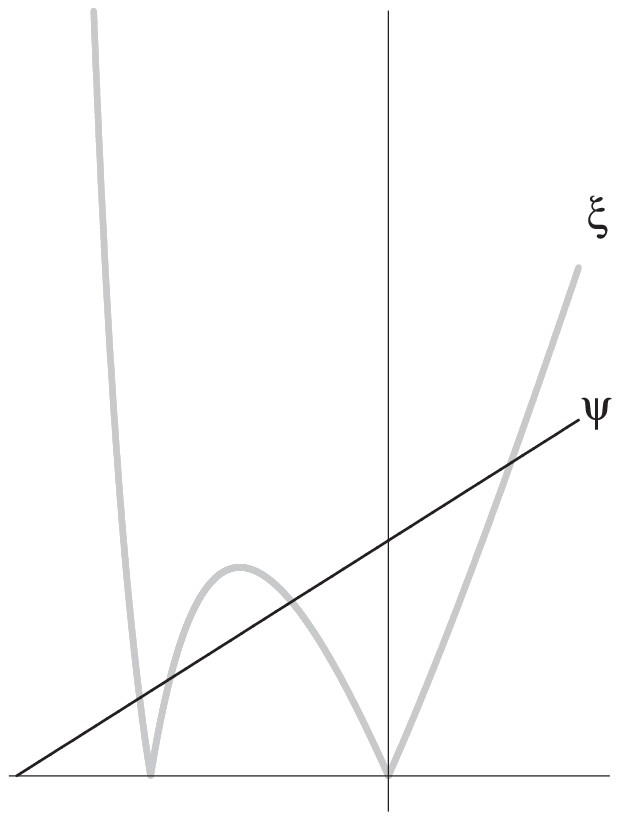} \\
(a) & (b) \\
\end{tabular}
\caption{ (a) $\psi$, $\xi$, and $Z$ plotted as a function of $y$ 
for a semiclassical black hole. The inner horizon of the classical
solution is replaced by a singularity where $\psi$, $\xi$, and $Z$ 
all approach zero. 
(b)~A corresponding plot for a classical black hole solution, 
repeated from Figure \ref{fig:classical} for comparison. }
\label{fig:psixi}
\end{figure}
Typical numerical results for massless matter are shown in 
Figure \ref{fig:psixi}a. We have also numerically integrated the 
equations with a non-vanishing fermion mass $m$ and find
the same qualitative behavior for small $m$. For large values
of $m$ the bosonic theory is strongly coupled and our
semiclassical equations can no longer be trusted.

The scalar fields $\psi$ and $Z$ extend smoothly through the
horizon at $y=0$, while $\xi$ goes to zero there and $\dot\xi$ 
changes sign, as in the classical theory. Away from the horizon
we see important departures from the classical solution.
We will discuss the exterior region below but let us first 
consider the interior of the black hole where the semiclassical
solution is dramatically different from the classical geometry.

\subsection{Black hole region}

In the classical black hole solution 
(\ref{classpsi})--(\ref{classxi}) the area function $\psi$ goes
linearly to zero with negative $y$ inside the event horizon.
The falloff of $\psi$ is more rapid in the corresponding 
semiclassical solution. 
This can be traced to pair-creation of charged fermions in the
black hole interior, which causes the amplitude of the
bosonized matter field to decrease as we go deeper into the
black hole.  From the constraint equation (\ref{staticconstraint})
we see that $\psi$ is a concave function and any variation in 
$Z$ serves to focus it towards zero.
The conformal factor of the metric is contained in the $\xi$ field.
Inside the event horizon at $y<0$ the left hand side of 
(\ref{staticxieq}) changes sign and as a result the $\xi$ field
is also concave in this region. Initially, $\xi$ grows away from
zero at the horizon as we go towards negative $y$ but eventually
it turns over and approaches zero again at a finite negative 
value $y=-y_0$.

\begin{figure}
\begin{tabular}{c@{\hspace{0.2cm}}c}
\psfrag{y}{$y$}
\psfrag{y0}{$-y_0$}
\psfrag{Z}{$Z$}
\includegraphics[width=3.8cm]{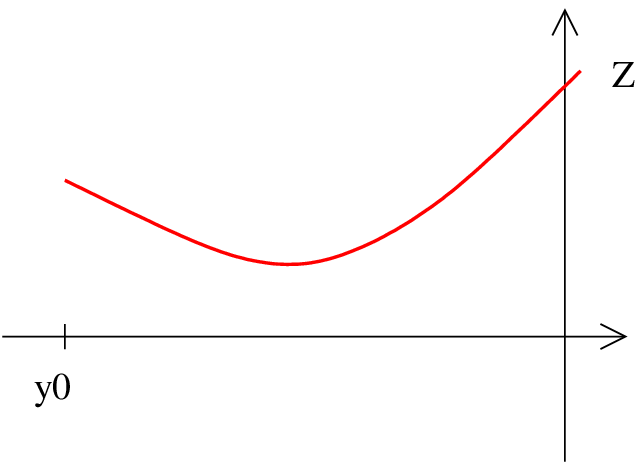} & 
\psfrag{y}{$y$}
\psfrag{y0}{$-y_0$}
\psfrag{Z}{$Z$}
\includegraphics[width=3.8cm]{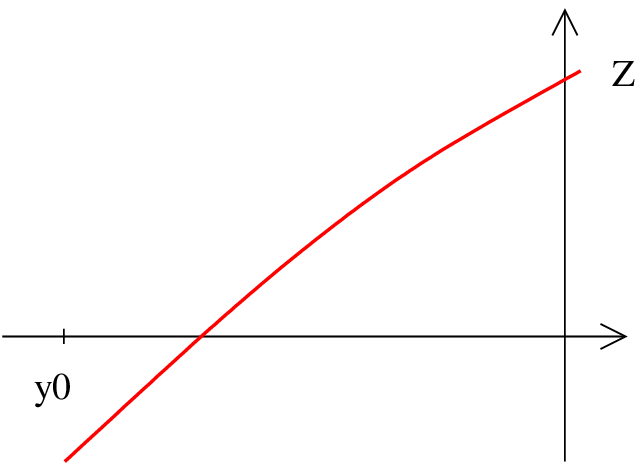} \\
(a) & (b)\\
\end{tabular}
\caption{Two cases in which the sign change of equation
(\ref{eq:logZ})  occurs.  
(a) The field $Z$ has a local minimum at some $y$
in the interval $-y_0<y<0$, or (b)~the field $Z$ goes through
zero at some $y$ in the same interval.}
\label{fig:signZ}
\end{figure}

This second zero of $\xi$ does not correspond to a smooth inner
horizon, as can be seen from the following argument. Finiteness
of $\ddot{\psi}$ and $\ddot{Z}$ at $y=-y_0$ would require
\begin{eqnarray}
\label{innerdatapsi}
\dot \psi(-y_0) &=&- \frac{1}{\dot \xi(-y_0)}\left(
4-\frac{Z(-y_0)^2}{2\pi\psi(-y_0)^2 } \right), \\
\label{innerdataz}
\dot Z(-y_0) &=& -\frac{1}{\dot \xi(-y_0)}\left(
\frac{Z(-y_0)}{\pi\psi(-y_0)^2 } \right),
\end{eqnarray}
where we have set the fermion mass $m$ to zero for simplicity.
Since $\dot \xi(-y_0)>0$ as $y \rightarrow -y_0$ it follows from 
equation (\ref{innerdataz}) that
\begin{equation}
\label{eq:logZ}
\frac{d}{dy}\log\vert Z\vert < 0, 
\qquad\textrm{as} \qquad y \rightarrow -y_0 ,
\end{equation}
while the same quantity is positive at $y=0$. This sign change
can occur in one of two ways, shown in Figure~\ref{fig:signZ}. 
Assuming $Z(0)>0$ (a parallel argument can be given for 
$Z(0)<0$) we either have a local minimum of $Z$ at some $y$ in
the interval $-y_0<y<0$, as in Figure~\ref{fig:signZ}a, or $Z$ goes 
through zero at some $y$ in the same interval, as in 
Figure~\ref{fig:signZ}b.
A local minimum for $Z$ is easily ruled out. Inserting 
$\dot Z=0$ into equation (\ref{staticzeqinside}) with $Z>0$ we
find that $\ddot Z<0$, which corresponds to a local maximum
rather than a minimum. The other possibility, {\it i.e.} $Z$
going through zero, can be eliminated on physical grounds, 
because it would correspond to the charge inside a given
location changing sign. Screening due to pair production 
will reduce the charge as one goes deeper into the black hole
but cannot change its sign. We cannot rule out that non-linear
effects in a strong coupling region close to a curvature 
singularity could reverse the sign of $Z$, but the above argument
does rule out the possibility of having an inner horizon with
macroscopic area.

We conclude from this that the solution will not be smooth
as $\xi\rightarrow 0$ inside the black hole and this is borne
out by  our numerical calculations. We find that all the fields 
$\xi$, $\psi$ and $Z$ are simultaneously driven to zero while 
their derivatives become large.  The Ricci scalar
\begin{equation}
R = -\xi \ddot \psi - \dot\xi\dot\psi 
      + \ddot \xi \psi + \frac{\xi \dot \psi^2}{\psi^2},
\label{staticricci}
\end{equation} 
evaluated for the numerical solution, increases rapidly as we
approach the singular point. This indicates that the classical
inner horizon is replaced by a curvature singularity at the 
semiclassical level.  The numerical solution breaks down
before an actual singularity is reached, but as far as the
numerical evidence goes, it indicates that the singularity is
spacelike.  In section \ref{sec:dynamical} we will present 
numerical evidence that the singularity formed in dynamical 
gravitational collapse in this model is also spacelike.  

We have found a two-parameter family of singular scaling 
solutions, which is a candidate for the final approach to the 
singularity,
\begin{eqnarray}
\label{psisingular}
\psi(y)
&=& A (y+y_0) [-\log(y+y_0)]^a ,
\\ \label{xisingular}
\xi(y)
&=& \frac{4}{\pi A^2} [-\log(y+y_0)]^{-2a} ,
\\ \label{zsingular}
Z(y)
&=& \pm\sqrt{8aA}(y+y_0)^{1/2} [-\log(y+y_0)]^{(a-1)/2} ,
\end{eqnarray}
Here $y=-y_0$ is the coordinate location of the singularity,
which depends on the global coordinate scale set by $\dot \xi(0)$ 
in equations (\ref{initialdatapsi}) and (\ref{initialdataz}). 
We emphasize that this singular solution is at best asymptotic 
to the true solutions near the singularity, since it only equates 
the most singular terms in the semiclassical equations, leaving 
behind terms that are sub-leading but nevertheless divergent. 
The neglected sub-leading terms are only logarithmically 
suppressed compared to the leading terms and this means that we 
are unable to match our numerical solution onto the proposed 
scaling solution.  A successful match would either involve 
following the numerical solution extremely close to the 
singularity, far beyond presently attainable numerical precision, 
or working out several subleading orders in the scaling solution, 
which is beyond our analytic perseverance. In the absence of 
successful matching we can only tentatively claim that
our scaling solution correctly describes the geometry near
the singularity, but let us nevertheless investigate some of 
its properties.

The Ricci scalar (\ref{staticricci}) is easily seen to diverge
as $y\to -y_0$ in the scaling solution 
(\ref{psisingular})--(\ref{zsingular}) so the singular point 
represents a true curvature singularity. The singularity is in 
fact strong, in the sense that not only does the curvature itself
blow up there but also its integral along a timelike geodesic
approaching $y=-y_0$.  This indicates that both the tidal
force acting on an extended observer and the integrated
tidal force will diverge as the singularity is approached.

The conformal factor of the metric, which is given by the
ratio $\xi/\psi$, diverges as $y\to -y_0$. A singularity
described by equations (\ref{psisingular})--(\ref{zsingular})
is therefore spacelike rather than null. The area function 
$\psi$ goes to zero as the singularity is approached in both the 
numerical solution and our proposed scaling solution. This means 
that the gravitational sector becomes infinitely strongly 
coupled at the curvature singularity.

\subsection{Exterior region}

The spacelike singularity encountered in the black hole interior 
is the most important feature of our semiclassical solutions but
it is useful to consider also the region far away from the black 
hole. Understanding the asymptotic behavior of the solutions 
provides a check on our formalism.
In this region the coupling strength $e^{\phi}$ goes to zero 
and the gravitational fields $\psi$ and $\xi$ should approach
classical behavior.  Both $\psi$ and $\xi$ grow linearly with
$y$ in the classical solution (\ref{classpsi})--(\ref{classxi}).
The corresponding semiclassical fields also grow linearly
at large $y$ but have additional logarithmic terms.
The static equations (\ref{staticpsieq})--(\ref{staticconstraint})
are solved to leading order at large $y$ by
\begin{eqnarray}
\label{asextz}
 Z(y) &=& Z_0 \log(\alpha y) + \cdots \\ 
\label{asextpsi}
\psi(y) &=& \alpha y + \frac{Z_0^2}{2} \log(\alpha y) + \cdots \\
\label{asextxi}
\xi(y) &=&   \frac{4}{\alpha} y + \xi_0 \log(\alpha y) + \cdots
\end{eqnarray}
where the $\cdots$ denote terms that are constant or vanish
in the limit. 

As a result of the logarithmic growth of $Z$ there is a 
non-vanishing energy density in matter in the asymptotic region 
and, via equation (\ref{bosoncurrent}), a finite electric charge 
density also.  The ADM total mass and total charge of these
black hole geometries are therefore infinite. At first sight, one 
might expect that an infinite amount of matter energy will collapse
the geometry, but the gravitational coupling goes to zero in 
the asymptotic region so the matter in fact decouples from the
gravitational sector.  In two spacetime dimensions we can 
interpret a non-vanishing asymptotic energy density of 
massless matter in a static solution in terms of balanced
incoming and outgoing energy fluxes carried by particles
in the limit of zero momentum \cite{Peet:1993vb}. In our case,
the elementary fermions carry electric charge so there
is also a balance between outgoing and incoming electric flux.

The outgoing flux is due to pair-production in the electric field
outside the event horizon of the black hole. Particles carrying 
charge of opposite sign to that of the black hole are electrically
attracted to the hole while same sign charges are repelled. 
This leads to a net flow of charge and energy away from the black 
hole but since we are considering static solutions this flow 
is balanced by an equal influx from an external source at infinity. 
This is reminiscent of the so-called quantum black hole solutions
of \cite{Birnir:1992by} and the interpretation is similar, except 
in that case the outgoing energy flux at infinity is in the form 
of Hawking radiation rather than charged particles formed by 
pair-production. 
 
\begin{figure}
\psfrag{eps}{$\varepsilon$}
\psfrag{f}{$f(0)/\sqrt{8}$}
\psfrag{psi}{$\psi(0)$}
\includegraphics[width=0.8\columnwidth]{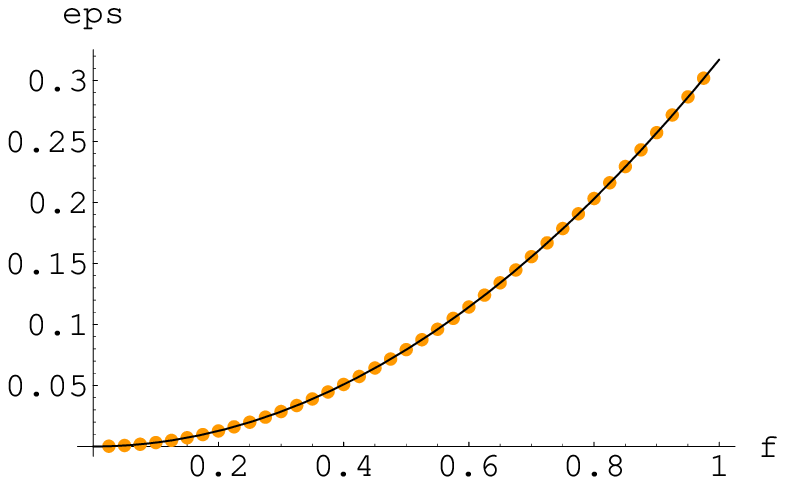} 
\hspace{0.6in} (a)  \\
\includegraphics[width=0.8\columnwidth]{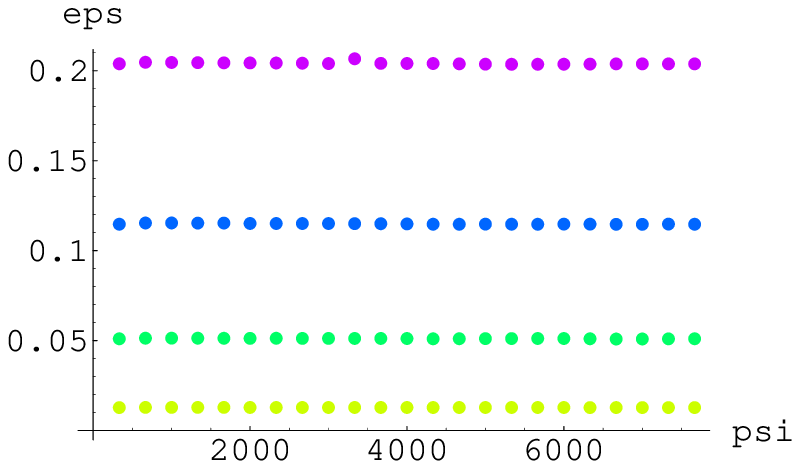} 
\hspace{0.6in} (b)  \\
\caption{
(a) The asymptotic energy density $\varepsilon$ as a function
of the electric field at the horizon $f(0)$.
(b)~The asymptotic energy density $\varepsilon$ as a function
of $\psi(0)$ for four values of 
$f(0)/\sqrt{8} = 0.2, 0.4, 0.6, 0.8$.
}
\label{fig:energydensity}
\end{figure}

The asymptotic matter energy density, measured with respect to an 
asymptotically Minkowskian coordinate frame, is given by
\begin{equation}
\label{energydensity}
\varepsilon=\lim_{y\rightarrow\infty}\frac{1}{2}\psi\xi 
\left(\frac{dZ}{dy}\right)^2,
\end{equation}
which reduces to $\varepsilon = 2Z_0^2$ for the semiclassical
fields in equations (\ref{asextz})--(\ref{asextxi}). 
The value of $\varepsilon$ for a given semiclassical black
hole is easily obtained from the numerical solution 
and the results are plotted in Figure~\ref{fig:energydensity}.
The data clearly show that the energy density depends on
the electric field at the black hole horizon but is
independent of the overall size of the black hole for a fixed
electric field.
We are using equation  (\ref{zforgauge}) to define the electric
field strength at the horizon as~\footnote{Strictly speaking this is 
$-f(0)$ but we choose to absorb the
minus sign into the definition of $f(0)$ in order to avoid 
cluttering our equations with minus signs.} 
\begin{equation}
f(0) = \frac{Z(0)}{\sqrt{\pi} \psi(0)} .
\end{equation}

For massless fermions the exponential suppression of Schwinger
pair-production is absent and the pair production rate per proper
volume depends on the electric field strength as a power law.
In two spacetime dimensions this dependence is linear but
this is only part of the story.
Particles are created at every distance from the black hole, 
although most of them come from near the horizon where the electric
field is strongest, and those particles that carry the same sign 
charge as the black hole are then accelerated out to infinity by 
the electric field. These particles, along with a corresponding 
influx of energetic particles to maintain the static nature of the 
solution, make up the asymptotic matter energy density 
$\varepsilon$. The dots in Figure~\ref{fig:energydensity}(a) are 
results from numerical integrations while the solid line is given 
by a curve $\varepsilon \sim f(0)^2$. The close fit shows that the 
energy density at infinity goes like the electric field strength at 
the horizon squared.

\subsection{Maximally charged black holes}

At the classical level the extremal black hole of mass $M$ carries
the maximum possible charge for a black hole of that mass.
Adding more charge, keeping the mass fixed, results in a naked
singularity instead of a black hole.  The geometry of an extremal
black hole is qualitatively different from that of a non-extremal 
black hole in the classical theory.  The two horizons have merged 
into a double horizon which is located at infinite proper distance 
from fiducial observers outside the black hole.  This picture is 
modified in the semiclassical theory with $m=0$ where we find that 
there is again a maximum charge that can by carried by a black hole 
of a given mass but the corresponding geometry is not extremal.

The absence of extremal black holes is explained by the screening
effect of charged pairs produced in the electric field outside the
black hole.  Recall from equation (\ref{staticconstraint}) that 
$\psi$ is a concave function of $y$ and in order for the spacetime 
to be asymptotically flat $\psi$ must remain a growing function of 
$y$ in the asymptotic region $y\to\infty$. If the charge-to-mass 
ratio of the semiclassical black hole becomes too large the 
solution collapses to $\psi = 0$ at a finite distance outside the 
horizon and the geometry is no longer that of a black hole.
This is most conveniently analyzed in terms of the electric field at
the black hole horizon, $f(0)$, which is bounded from above in the 
classical limit by the field of an extremal black hole, 
$f(0) \le \sqrt{8}$.  A close look at our numerical solutions 
reveals that $\psi$ will collapse far away from the black hole
unless the electric field at the horizon satisfies
\begin{equation}
\label{maxf} 
\frac{f(0)}{\sqrt{8}} < 1 - \frac{1}{8\pi \psi(0)} 
+ {\cal O}\left(\frac{1}{\psi(0)^2}\right),
\end{equation}
which is lower than the classical maximum.  By looking at equation
(\ref{initialdatapsi}) for $m=0$ (and therefore $V(Z) = 0$) we see 
that $\dot\psi(0) > 0$ for all $f(0) < \sqrt{8}$ but if the 
electric field at the horizon is in the range 
\begin{equation}
1 - \frac{1}{8\pi \psi(0)} 
+ {\cal O}\left(\frac{1}{\psi(0)^2}\right)
< \frac{f(0)}{\sqrt{8}} < 1
\end{equation}
for large $\psi(0)$, then $\dot \psi$ changes sign at some $y>0$ 
and the solution collapses to $\psi = 0$.

The interior geometry of any semiclassical black hole, for which 
the electric field at the horizon is below the maximal value in 
equation (\ref{maxf}), looks like the typical solution in 
Figure~\ref{fig:classical}(b). In particular there is no sign of 
the double horizon of the classical extremal black hole in 
Figure~\ref{fig:psixi}(b).  We conclude that the energy density of 
pair-created particles outside a two-dimensional charged black hole 
will collapse the geometry before the extremal limit is reached.

\section{Dynamical collapse}
\label{sec:dynamical}

In this section, we study the problem of dynamical collapse of a
charged matter distribution by solving the semiclassical equations
(\ref{psieq})--(\ref{constraints}) numerically. It is the
internal structure of the resulting black hole that is of prime 
interest to us. We must therefore choose our coordinate system 
carefully (it should penetrate horizons) and select a numerical 
method which avoids propagating information faster than true 
characteristic speed of the equations. We employ a grid based on 
double-null coordinates $(u,v)$ and discretize the second-order 
equations of motion (\ref{psieq})--(\ref{zeq}) using a variation 
of a leapfrog algorithm.

\begin{figure}
\centerline{\epsfig{file=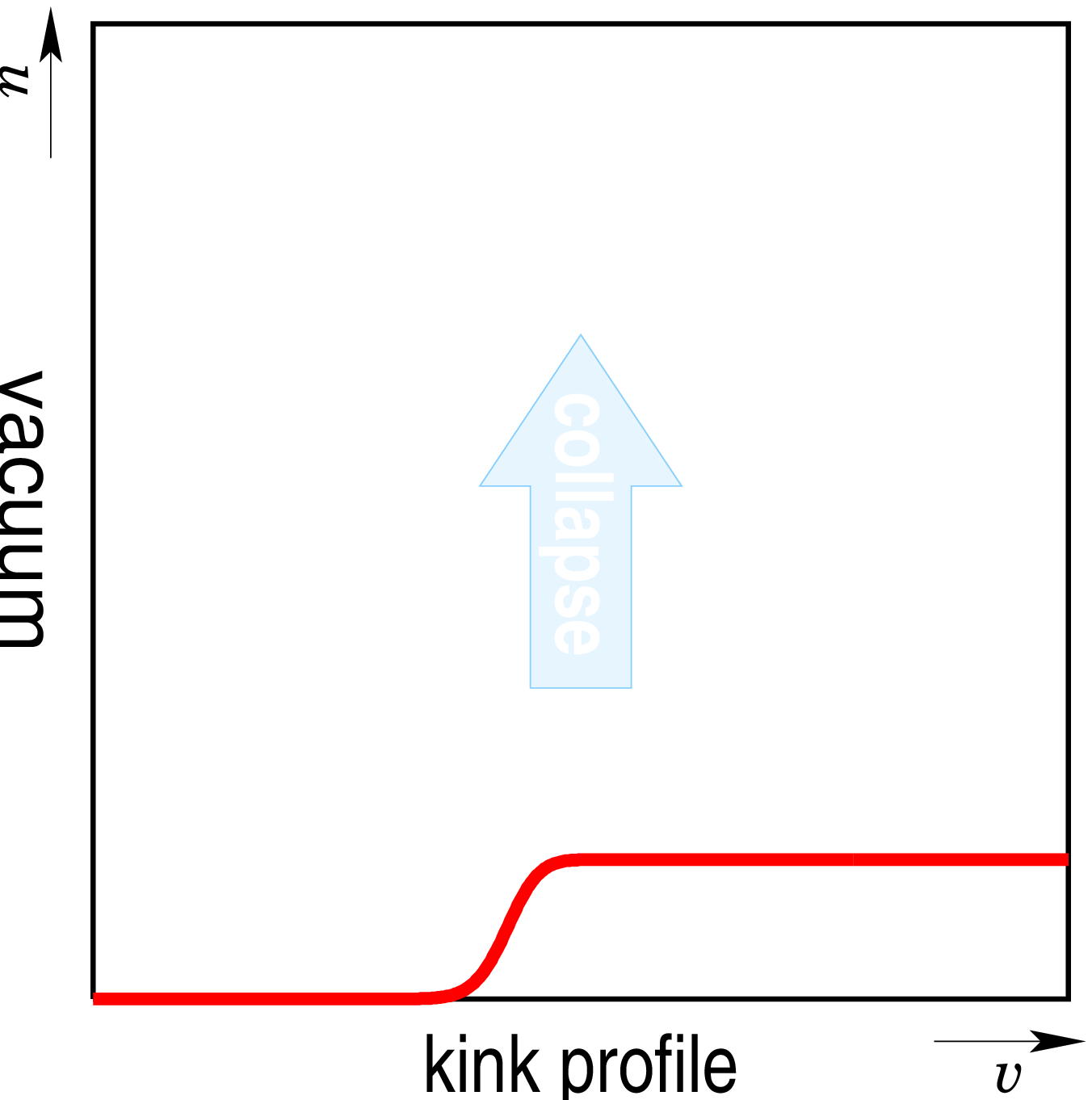, height=2.0in, angle=45}}
\vspace{-24pt}
\caption{Initial data for dynamical collapse.}
\label{fig:cauchy}
\end{figure}

The initial data for the null Cauchy problem is specified by 
providing the values of the functions $\psi$, $\theta$, and $Z$ 
on a double-null wedge, as shown in Figure~\ref{fig:cauchy}. 
Only one of the three functions is physical and we have chosen
this to be the bosonized matter field $Z$. The area function 
$\psi$ can be fixed by a choice of null coordinate parametrization 
on the initial wedge and then the remaining function $\theta$ is 
determined by solving the constraint equation (\ref{constraints}).

The physical initial condition is the initial incoming charge 
distribution $Z$, which we take to be a smooth kink profile 
\begin{equation}
  Z(0,v) = \frac{Q}{2} \left[\tanh\left(w \tan \pi 
\big({\textstyle \frac{v-v_1}{v_2-v_1}- \frac{1}{2}}\big)\right) 
+ 1\right]
\label{initialprofile}
\end{equation}
collapsing into a previously empty spacetime with $Z(u,0)=0$, as
illustrated by Figure~\ref{fig:cauchy}. The kink in the profile is 
localized between $v_1$ and $v_2$. It carries a total charge $Q$ 
and an energy density determined by its gradient squared, so the 
total mass scales roughly as $Q^2w$.

The null coordinate choice is effectively given by a choice of 
$\psi$ profile on the initial wedge. As the vacuum solution takes a 
particularly simple form in a gauge
\begin{equation}
  \psi(u,v) = - (u-u_0)(v-v_0),
  \hspace{1em}
  \theta(u,v) = 0,
\end{equation}
one is tempted to take $\psi(u,0)$ and $\psi(0,v)$ as linear
functions of $u$ and $v$ respectively. However, in order to cover 
the entire spacetime by a finite grid, we compactify the $v$
coordinate by taking, for example, 
$\psi(0,v) = \psi_0 + \alpha \tan k v$. Here $v$ runs over the range
$[0,1]$ and the parameter $k< \pi/2$ provides a regulator
for $\psi$ as $v\rightarrow 1$.
On the other hand, we leave $\psi(u,0)$ linear in $u$. With incoming
matter we expect a curvature singularity to form at $\psi=0$ and 
since $\psi(u,0)$ has a zero at a finite advanced time $u=u_0$ 
there is no need to compactify the $u$ coordinate.

\begin{figure}
\centerline{\epsfig{file=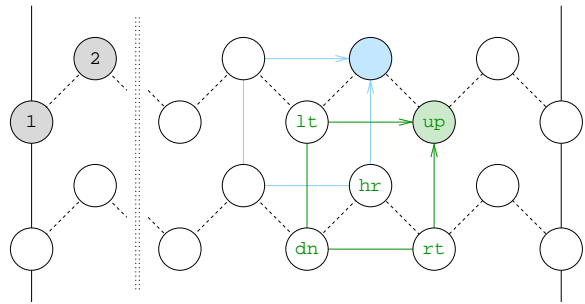, width=3in}}\vspace{-6pt}
\caption{The numerical evolution scheme is a a staggered grid 
leapfrog.}
\label{fig:grid}
\end{figure}

The remaining function $\theta$ is determined by fourth order 
fixed step Runge-Kutta integration of the constraint equation 
(\ref{constraints}) on the initial slice $u=0$. We also integrate 
$\partial_u$ and $\partial_{uu}$ derivatives, which are then used 
to jump-start the two-dimensional evolution.

The numerical evolution scheme is a leapfrog algorithm on a 
staggered grid, which is illustrated by Figure~\ref{fig:grid}. 
The values of functions at a given $u$ are stored in an array, 
and are propagated to the next time step $u+\epsilon$ using 
the equations of motion, except for the leftmost values, which 
are filled in using the vacuum initial conditions at $v=0$. 
The differential operators on the left hand sides of equations 
(\ref{psieq})--(\ref{zeq}) are discretized as
\begin{equation}\label{discrete}
-\partial_u\partial_v X = \frac{1}{\epsilon^2}
(X_{\text{lt}} + X_{\text{rt}} - X_{\text{dn}} - X_{\text{up}}),
\end{equation}
which gives a second order accurate expression in the center of 
the cell formed by $X_{\text{lt}}$, $X_{\text{up}}$, 
$X_{\text{dn}}$, and $X_{\text{rt}}$. This is why the staggered 
grid was chosen -- it provides the values of the fields at 
the correct location for evaluation of the right hand sides. 
One can think of the staggered grid as a square grid rotated by 
forty five degrees. Then the discretization (\ref{discrete}) is 
readily recognized as the usual leapfrog discretization of the 
wave operator $-\partial_t^2 + \partial_x^2$.

\begin{figure*}
  \begin{center}
  \begin{tabular}{cc}
    (a) $Z(u,v)$, $m=0$ & (b) $Z(u,v)$, $m=0.05$\\
    \epsfig{file=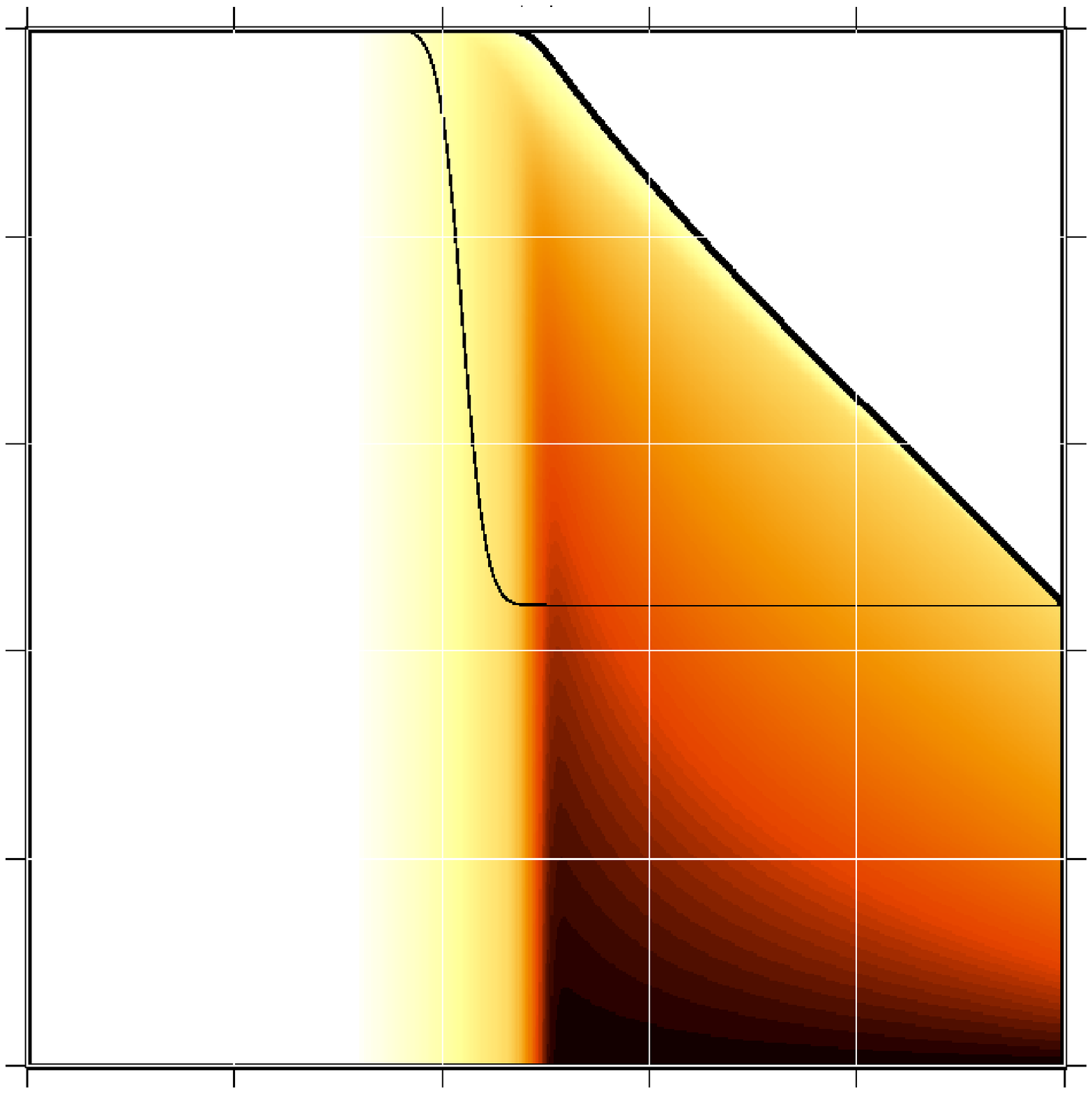, height=2.5in, angle=45} &
    \epsfig{file=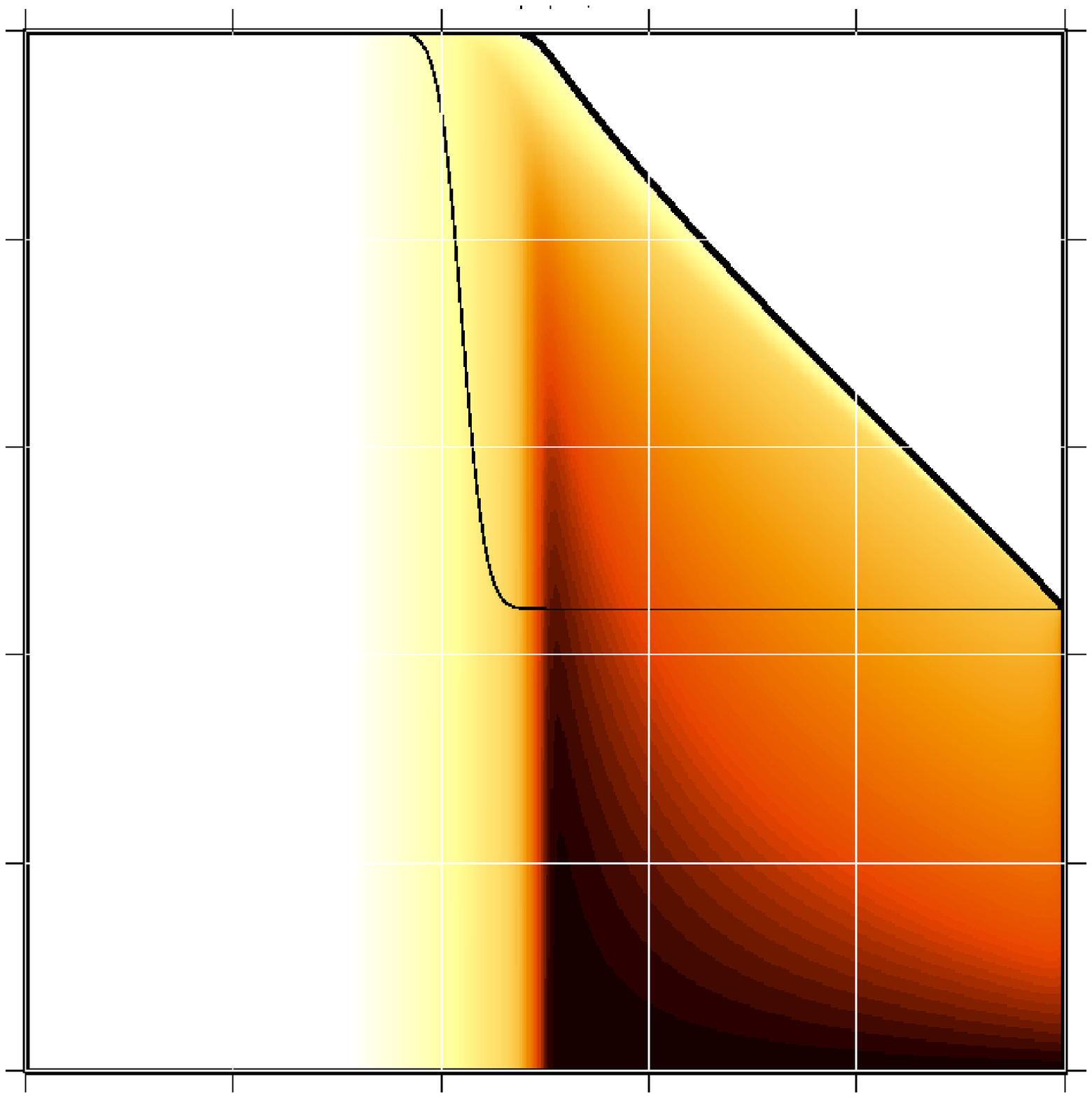, height=2.5in, angle=45} \medskip\\
    (c) $\psi(u,v)$, $m=0$ & (d) $R(u,v)$, $m=0$\\
    \epsfig{file=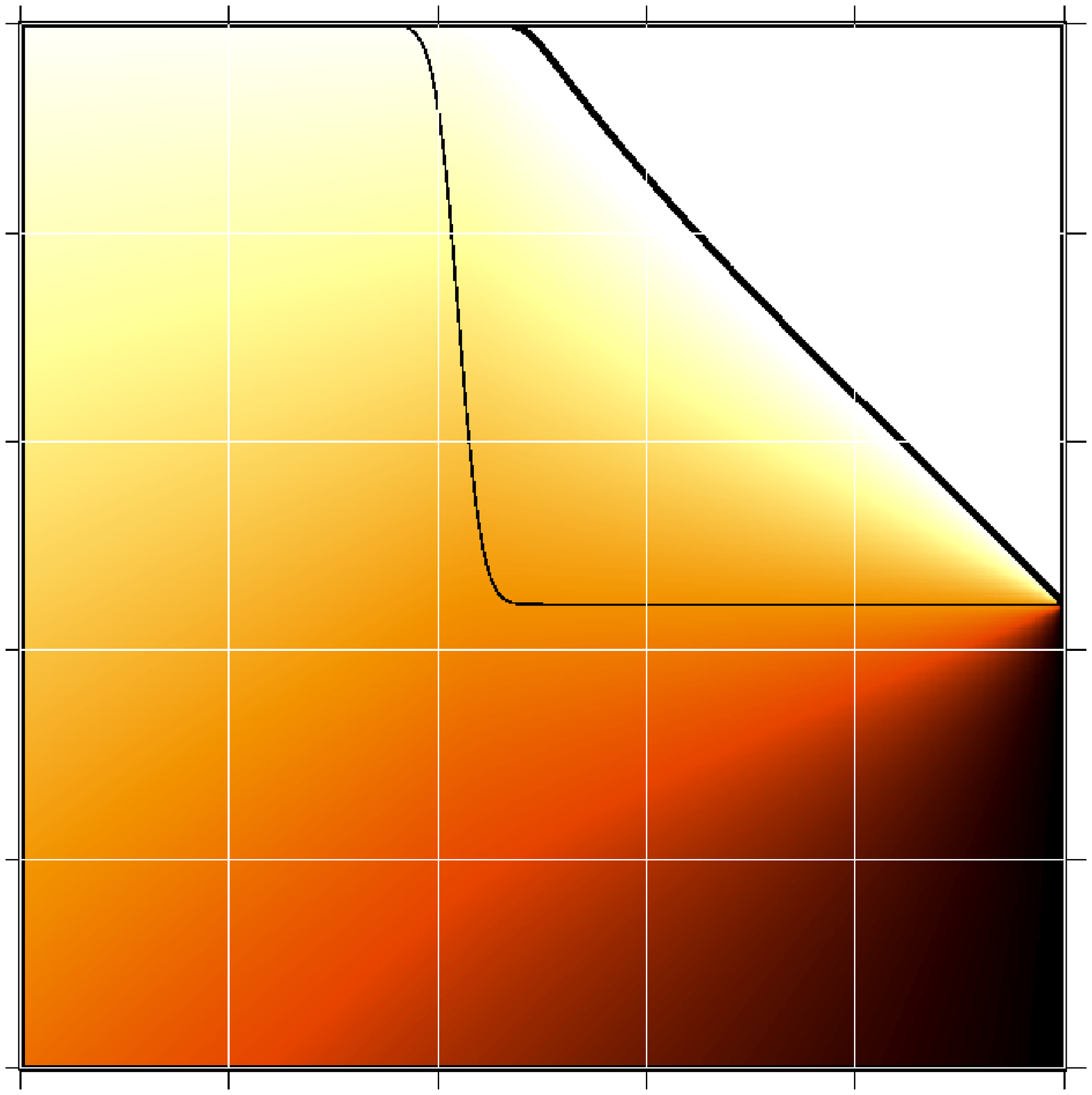, height=2.5in, angle=45} &
    \epsfig{file=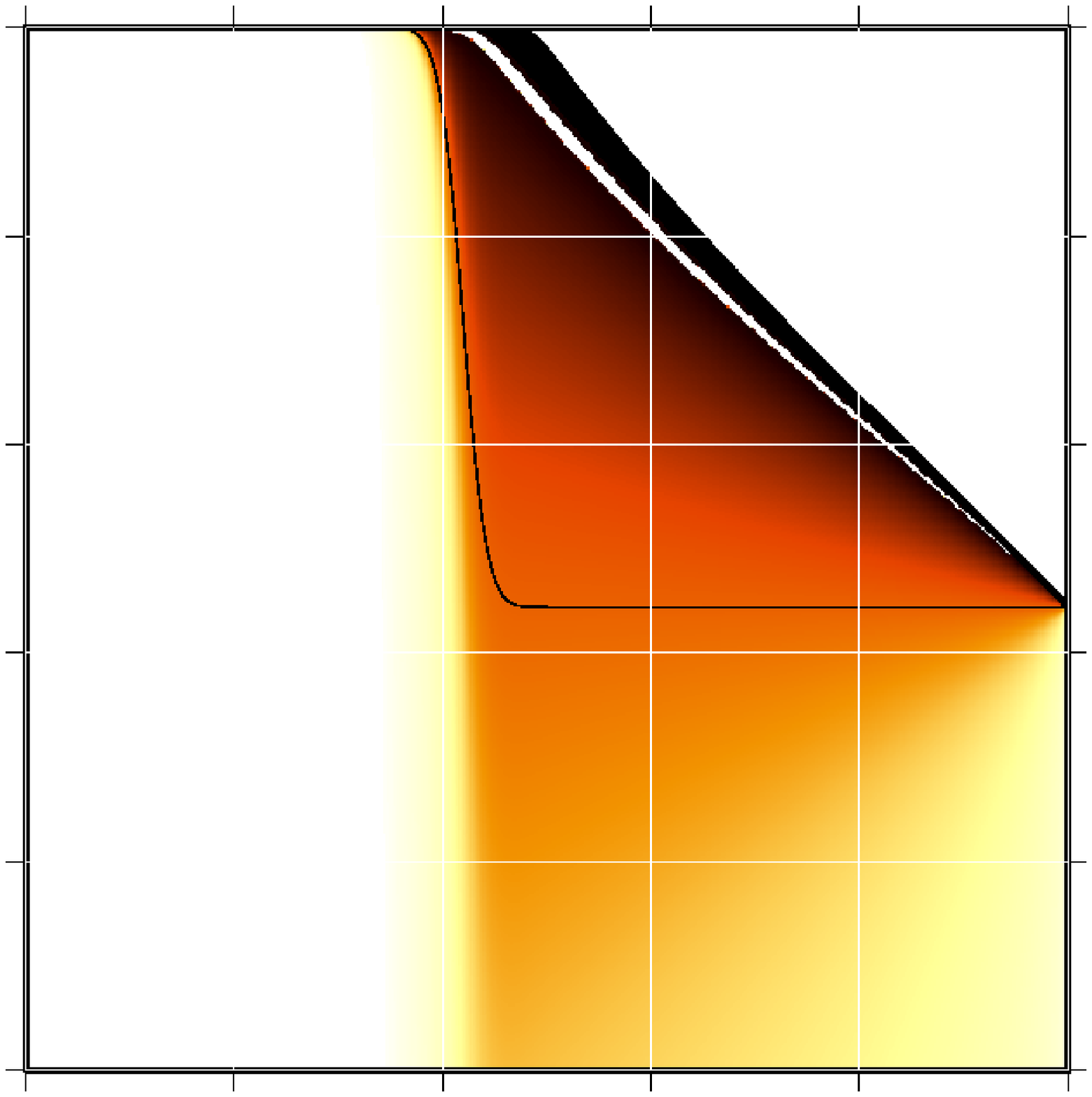, height=2.5in, angle=45} \\
  \end{tabular}
  \end{center}
  \vspace{-16pt}
  \caption{
Density plots of the charge distribution $Z(u,v)$ for the collapse of massless 
(a) and massive (b) fermions. The density plots of (c) dilaton $\psi(u,v)$ and 
(d) scalar curvature $R(u,v)$ are for collapse of massless fermions. The 
corresponding plots for collapse of massive fermions are not substantially 
different, and are not provided here.  The singularity is shown by a thick 
black line, while the thin line indicates the apparent horizon. The white stripe
just below the singularity in (d) is region of negative curvature.
}
\label{results}
\end{figure*}

\subsection{Numerical results}
We now present results obtained from numerical evolution using
the above algorithm and initial data of the type shown in 
Figure~\ref{fig:cauchy}. The results are exhibited as density
plots of the various fields with shading intensity giving the 
value of the field in question. To facilitate the interpretation
of the numerical data we indicate curvature singularities in the 
plots by thick black curves and apparent horizons by thin black curves. 
The local condition for a future trapped event is that the area function 
be decreasing along both future null directions, 
\begin{equation}
\partial_\pm \psi < 0 .
\end{equation}
The apparent horizon is located at the boundary of such a region, 
{\it i.e.} where 
$\partial_+ \psi = 0$ or $\partial_- \psi = 0$ \cite{Russo:1992ht}.

Figure~\ref{results}(a) shows the charge distribution $Z(u,v)$ 
in a spacetime where matter in the form of massless fermions undergoes
gravitational collapse. Observe how $Z\rightarrow 0$ deep inside the black 
hole, which can be attributed to charge screening due to fermion pair
production. One can also see the (slow) discharge of a black hole by pair 
production at the horizon. The shading scheme is exaggerated to show this 
more clearly.  For comparison, Figure~\ref{results}(b) shows the collapse 
resulting from the same initial conditions, but for massive fermions. 
In this case, the charge penetrates deeper into the black hole before 
pair production can screen it efficiently. The geometry of the spacetime 
is not substantially different for the massless and massive cases
and is only shown for the former.  Figure~\ref{results}(c) depicts the 
dilaton field $\psi(u,v)$, while Figure~\ref{results}(d) shows the 
scalar curvature $R(u,v)$ with compressed shading to span a huge range.
From these plots one clearly sees the formation of a spacelike curvature
singularity in the black hole interior and there is no indication of
a null singularity or a Cauchy horizon. 

\begin{figure}
\centerline{\epsfig{file=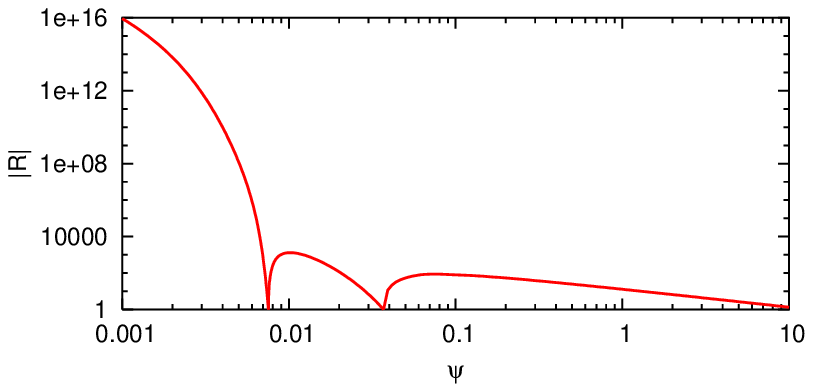, height=1.6in}}
\caption{Spacetime curvature along a constant $v$ profile that 
runs into the black hole singularity.}
\label{fig:logR}
\end{figure}

The white stripe preceding the singularity in the curvature plot in
Figure~\ref{results}(d) is a region of negative curvature, which the 
plot program treats as missing data. This oscillation in the spacetime 
curvature is also seen in Figure~\ref{fig:logR} which shows the
curvature along a profile at fixed retarded time, $v=$constant.
The figure is a log-log plot of $\vert R\vert$ against $\psi$ along the 
profile and the middle hump is in the region of negative curvature.

\subsection{Scaling behavior}
Rather remarkable critical behavior occurs at the onset of black hole 
formation in gravitational collapse in classical Einstein gravity coupled
to a massless scalar field \cite{Choptuik:1992jv}. A sufficiently weak
ingoing s-wave pulse reflects from the origin without creating a black 
hole but above a certain critical threshold, as the amplitude of the
pulse is increased, a black hole will form. Near this threshold the
black hole mass obeys a scaling law,
\begin{equation}
\log M = \gamma \log \delta + O(\delta^0) ,
\end{equation}
where $\delta$ parametrizes the distance from the threshold in the
initial data and $\gamma \approx 0.37$  is a scaling exponent 
obtained from numerical data \cite{Choptuik:1992jv}. 

\begin{figure}
\centerline{\epsfig{file=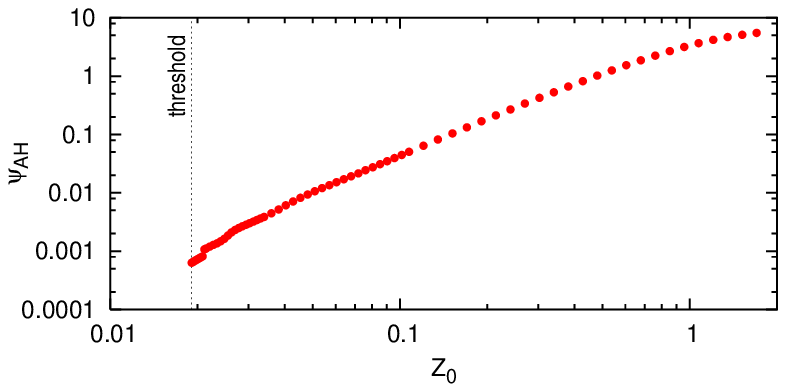, height=1.6in}}
\caption{Scaling behavior of the area of the apparent horizon.}
\label{fig:scaling1}
\centerline{\epsfig{file=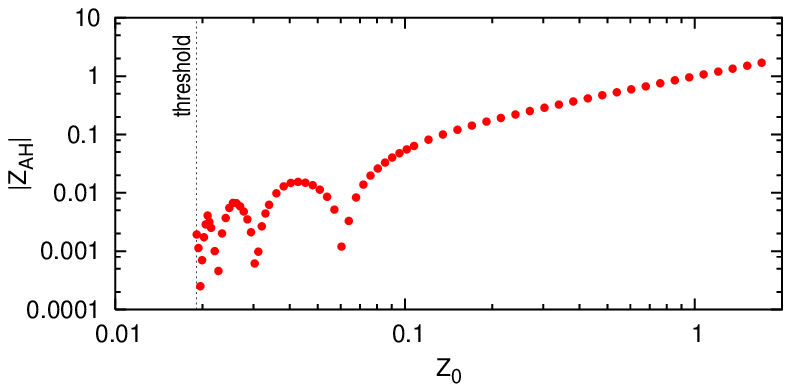, height=1.6in}}
\caption{The absolute value of $Z_{\text{AH}}$, the bosonized
matter field at the apparent horizon, in the scaling regime. 
The sign of $Z_{\text{AH}}$ changes between adjacent humps in the
graph.}
\label{fig:scaling2}
\end{figure}

We have looked for analogous scaling behavior in the formation
of charged black holes in our semiclassical model. 
Figure~\ref{fig:scaling1} shows the area of the apparent horizon
that forms in gravitational collapse of charged matter as a 
function of the amplitude of the incoming kink profile.  A close 
look at the numerical data indicates that there is indeed a 
threshold for black hole formation, which we find by bisection.
The data also suggest that 
black hole formation turns on at finite size, which is reminiscent 
of the Type~I critical behaviour found in gravitational collapse 
of a Yang-Mills field in Einstein gravity \cite{Choptuik:1999gh}.
For a super-critical collapse, the size of the black hole can be
fitted by a power law
\begin{equation}
\psi_{\text{AH}} \propto (Z_0-Z_*)^{\gamma_Z} \,,
\end{equation}
where $Z_0$ is the amplitude of the initial kink,
and the scaling exponent is
\begin{equation}
\gamma_Z = 1.85 \pm 0.01 \,.
\end{equation}
We have also considered a family of initial data where the
width parameter $w$ of the kink profile in equation 
(\ref{initialprofile}) is varied rather than its height.
We again find scaling behavior,
\begin{equation}
\psi_{\text{AH}} \propto (w-w_*)^{\gamma_w} \,,
\end{equation}
but with a different scaling exponent
\begin{equation}
\gamma_w = 0.71 \pm 0.03 \,,
\end{equation}
so it appears that the scaling exponent is not universal.
We should note that the best-fitting offset values $Z_*$ and $w_*$
are substantially less than the actual threshold values.

By equation (\ref{zforgauge}) the bosonized matter field $Z$ 
determines the amount of charge to the left of, or inside, a given 
location. It is therefore straightforward to consider the initial
black hole charge, which we define as the charge inside the 
apparent horizon when the black hole forms, for initial data 
in the scaling regime.  The results, presented in 
Figure~\ref{fig:scaling2}, show more complicated behavior than that of the black 
hole area. The black hole charge appears to show an overall power-law scaling,
but the sign of the charge starts oscillating and the power-law exponent steepens as the threshold is 
approached. It would be interesting to understand this behavior 
better.

\section{Discussion}

We have studied the geometry of charged black holes in the 
context of a 1+1-dimensional model of dilaton gravity with 
charged matter in the form of bosonized fermions.  At the 
classical level, the model has static black hole solutions 
with a global causal structure identical to the four-dimensional 
Reissner-Nordstr\"om solution. When the quantum effect of 
pair-production of charged particles is taken into account 
the classical inner horizon is replaced by a spacelike singularity
and the global structure of the spacetime geometry is that of an 
electrically neutral black hole.

This result is based on a combination of numerical and analytic
analysis of semiclassical equations of motion, both static and 
dynamical. The static solutions are being constantly fed by an 
external source to balance quantum mechanical black hole 
evaporation and discharge. The dynamical solutions, on the other 
hand, describe the collapse of charged matter into vacuum.

The strong form of the cosmic censor conjecture 
\cite{Penrose:1980ge} forbids 
naked singularities to be visible to any physical observers,
including ones who travel inside charged black holes. Our 
semiclassical results clearly support strong cosmic censorship, 
while its validity for charged black holes in classical gravity 
is a delicate issue \cite{Ori:1991,Dafermos:2003wr}.

We believe our two-dimensional model captures some of the 
essential physics of this problem while leaving out many of
the complications of the full higher-dimensional quantum
gravity problem. The model is certainly not without fault,
it has no propagating gravitons or photons and the electric
field of a black hole only depends on the charge to mass ratio 
of the black hole, but as far as we know there exists no 
systematic treatment of quantum effects, including pair-production
of charged particles, for higher-dimensional black holes.

Our results can be improved on in a number of ways.  For 
example, we have not included gravitational quantum effects 
in our two-dimensional model in this paper. There exists an 
extensive literature on semiclassical two-dimensional dilaton 
gravity, including the Hawking effect and its back-reaction on 
the geometry of neutral black holes. For reviews see 
\cite{Giddings:1994pj,Strominger:1994tn,Thorlacius:1994ip,
Grumiller:2002nm,Fabbri:2005mw}. 
We expect that gravitational back-reaction to Hawking radiation 
will not change our main conclusion that the singularity formed 
in the gravitational collapse of charged matter is spacelike. 
Subtleties involving boundary conditions in the strong coupling
region complicate the problem but a preliminary numerical 
study of charged black hole formation, with the combined effect 
of pair-production and electrically neutral Hawking emission 
included, confirms this expectation \cite{Kristjansson:2005}.
Quantum effects due to electrically neutral matter in a charged 
black hole spacetime have been considered by a number of 
authors, see for example \cite{Trivedi:1992vh,Frolov:1996hd,
Diba:2002hb,Zaslavskii:2004tv}. However, pair-production is not 
considered in these papers. 

One might also worry about the phenomenological relevance of 
charged black holes in general. It is after all very unlikely 
that black holes carrying macroscopic electric charge are 
found in Nature. Solutions that describe such objects do,
however, exist in the theories we use to describe Nature and 
they provide an important testbed for theoretical ideas. 
Furthermore, issues of Cauchy horizons and non-trivial 
global topology of spacetime also arise in the context
of rotating black holes, which presumably are the generic 
black holes of astrophysics.

\acknowledgements

This work was supported in part by the Institute for Theoretical 
Physics at Stanford University, by grants from the Icelandic 
Research Fund and the University of Iceland Research Fund, 
and by the European Community's Human Potential Programme 
under contract MRTN-CT-2004-005104 ``Constituents, 
fundamental forces and symmetries of the universe". 

\bibliography{qa-heimildir}

\end{document}